\newcommand{\sugg}[1]{{#1}}

\newcommand{\openone}{\leavevmode\hbox{\small1\normalsize\kern-.33em1}}
\newcommand{\matriz}[1]{#1}

\documentclass[10pt]{iopart}
\usepackage{iopams,bm,cite,color,graphicx,subfigure,braket,mathptmx,url,enumitem}
\newtheorem{thm}{Theorem}

\usepackage[colorlinks=true,linkcolor=blue,citecolor=blue,urlcolor =blue]{hyperref}

\eqnobysec

\begin{document}

\title{Efficient tomography with unknown detectors}

\author{L~Motka$^{1}$, M~Pa\'{u}r$^{1}$, J~\v{R}eh\'a\v{c}ek$^{1}$,
  Z~Hradil$^{1}$ and L~L~S\'{a}nchez-Soto$^{2,3}$}

\address{$^{1}$ Department of Optics, Palack\'y University,
  17. listopadu 12, 771 46 Olomouc, Czech~Republic}
 
\address{$^{2}$ Departamento de \'Optica, Facultad de F\'{\i}sica,
  Universidad Complutense, 28040~Madrid, Spain}

\address{$^{3}$ Max-Planck-Institut f\"ur die Physik des Lichts,
  Staudtstra{\ss}e 1, 91058 Erlangen, Germany}

\begin{abstract}
  We compare the two main techniques used for estimating the state of
  a physical system from unknown measurements: standard detector
  tomography and data-pattern tomography. Adopting linear inversion as
  a fair benchmark, we show that the difference between these two
  protocols can be traced back to the nonexistence of the
  reverse-order law for pseudoinverses. We capitalize on this fact to
  identify regimes where the data-pattern approach outperforms the
  standard one and \textit{vice versa}. We corroborate these
  conclusions with numerical simulations of relevant examples of 
  quantum state tomography.  

\bigskip

\end{abstract}

%\maketitle

\section{Introduction}
\label{sec:intro}

Tomography is a special case of an inverse problem: loosely speaking,
it aims at making quantitative inferences about a physical system from
carefully designed measurements~\cite{Groetsch:1993aa,
  Tarantola:2005aa,Aster:2013aa}.  Needless to say, its importance to
classical and quantum physics cannot be overestimated.

However, this tool requires an extremely precise tuning. Indeed, a
complete characterization of the measurement setup is of utmost
importance,  and this necessarily involves a calibration of it.  At the
classical level, this calibration is relatively simple, as the signals
are deterministic (apart from the presence of noise). However, the
quantum world reveals itself in terms of very weak signals that are
intrinsically random. This is why having efficient, precise and simple
quantum detector calibration methods is particularly challenging and
has been attracting increasing attention.

Quantum detector tomography is concerned precisely with a trustworthy
characterization of the measurement
device~\cite{Luis:1999yg,Fiurasek:2001dn,DAriano:2004oe}. Nowadays,
the standard approach is to reconstruct the action of the measurement
from the outcome statistics in response to a set of tomographically
complete certified input probes~\cite{Lundeen:2009sf,Zhang:2012fu}. To
date, this has been successfully applied to a variety of problems,
including avalanche photodiodes~\cite{DAuria:2011aa,Zhang:2012aa},
time-multiplexed photon-number-resolving
detectors~\cite{Feito:2009aa,Coldenstrodt:2009aa}, transition edge
sensors~\cite{Brida:2012mz} and superconducting nanowire
detectors~\cite{Akhlaghi:2011aa,Renema:2012aa,Renema:2014aa}.

This standard protocol becomes increasingly demanding as the number of
detector outcomes grows: it requires to acquire and analyze expanded
data sets, which soon becomes intractable with current experimental
and computational capacity. This has prompted the search for some
helpful shortcuts.  This is the case when one can describe the setup
with a few-parameter model, such as efficiency, dark-count rate, etc.
For example, with a twin-photon state, one can find the absolute value
of the detector efficiency~\cite{Klyshko:1980aa,Worsley:2009aa,
Avella:2011aa}.  In the same vein, entanglement also makes possible
self-testing~\cite{Yang:2014aa,Wu:2016aa,Chen:2016aa}. Trading
knowledge of probes for information about the measurement gives rise
to the concept of self-calibrating
tomography~\cite{Mogilevtsev:2009aa,Mogilevtsev:2010aa,
Branczyk:2012aa,Mogilevtsev:2012aa,Stark:2016aa}.

There is, however, the possibility of skipping the calibration stage
altogether. This constitutes the essence of the data-pattern
tomography~\cite{Rehacek:2010fk,Mogilevtsev:2013kl}, which has been
recently implemented with remarkable
success~\cite{Cooper:2013fq,Harder:2014aa,Altorio:2016aa}.  The idea
is to measure responses (the data patterns) for a set of known quantum
probe states and match them with the response obtained from the
unknown signal of interest. This approach is insensitive to
imperfections of the setup, for they are automatically accounted for
by the  data patterns.  In addition, it does not require any
assumption about the search subspace, which is naturally defined by
the choice of probe states.

It seems thus pertinent to appraise the fundamental differences
between these two strategies and assess their performance. This is
precisely the goal of this paper. To have a fair benchmarking, we
adopt linear inversion as it simplicity allows us to gain deeper
understanding into the problem. When applied to the same data, we find
that the distinctions between the standard and data-pattern techniques
can be tracked down to the nonexistence of the reverse-order law for
pseudoinverses. In particular, we argue that in minimal
tomography; i.e., when the number of  outcomes equals the
number of state parameters, the data-pattern tomography outperforms
the standard one, whereas the opposite is true for highly overcomplete
settings. We confirm and illustrate our theoretical findings with
simulations utilizing random and homodyne detections.

\section{Basics of linear inversion}
\label{sec:tomo}

The objective of tomography is to characterize a physical system,
here simply called the \textit{signal}, from  suitable measurements 
performed on the system. The signal depends on $n$ parameters encoded
in the vector $\bi{r}\in \mathbb{R}^{n}$.  The measurement is assumed
to have $m$ discrete outcomes, whose expected values are denoted as
$\bi{p} \in \mathbb{R}^{m}$.  In any inverse problem one has
to find the best model such that $\bi{p}=  A  (\bi{r} )$,
where $A$ describes the explicit relation between the observed data
$\bi{p}$ and the model parameters $\bi{r}$. It is called the forward
(or measurement) operator.  In many cases of interest the relation
between the signal and observations is linear, and we can write
\begin{equation}
  \label{eq:tomo}
  \bi{p}= \matriz{A}  \,  \bi{r} \, ,
\end{equation}
where $\matriz{A} $ is a unique $ m \times n$ measurement 
matrix.  

The measurements are invariably subject to some uncertainty, 
and this means that the collected data, we call $\bi{f}$, deviates
from the expected values $\bi{p}$. There are many contributors to the
uncertainty; they can be divided into systematic and
random components: the  systematic errors  consist primarily of
unmodeled physics, whereas the random errors  are, by definition,
those variations in the data that are not deterministically reproducible. 

The ultimate goal is to infer the unknown signal parameters $\bi{r}$
from the measured noisy data $\bi{f}$.  This amounts to providing a
sensible inversion of equation~(\ref{eq:tomo}).  At first sight, the
simplest way is a direct linear (\textsc{lin}) inversion; namely,
\begin{equation}
  \label{inversion}
  \bi{r} = \matriz{A}^{+} \, \bi{f},
\end{equation}
where the expected results $\bi{p}$ are replaced with real data
$\bi{f}$ and acted upon by the Moore-Penrose pseudoinverse of the
measurement matrix $\matriz{A}$, indicated by the superscript $+$
(see~\ref{sec:MP} for a short introduction to the pseudoinverse).

This \textsc{lin} estimator is also known as the ordinary least
squares (\textsc{ols}) estimator~\cite{Lawson:1974aa} and it is
unbiased and consistent. Under the Gauss-Markov
assumptions~\cite{Hallin:2006aa} it is also the best linear unbiased
estimator (\textsc{blue})~\cite{Kay:1993aa}.  The \textsc{lin} is also
the preferred linear estimator for small and medium sized data sets,
when a reliable estimation of the data covariances is not possible.
Although nonlinear techniques, such as maximum
likelihood~\cite{Millar:2011aa} or Bayesian
methods~\cite{Gelman:2013aa}, can be more efficient for general
correlated data, \textsc{lin} is more than enough for our purposes
here.
 
Let us illustrate the previous discussion with two examples.  
In an imaging system, $\bi{r}$ may describe the (sampled) true
object intensities, $\bi{p}$ the mean image intensities, $f_k$ one
particular image CCD scan, and the matrix $a_{jk}$ the discretized
convolution kernel expressing the intensity blurring due to the finite
aperture and/or optical aberrations.

Quantum tomography may serve as another example. Here, an unknown true
quantum state is  described by a $d \times d$ density matrix
$\varrho$, which requires $n= d^{2}-1$ independent parameters for its
complete specification.  In general,  the measurements
performed on the system are described by  positive operator-valued
measures (POVMs); which are a set of operators 
$\{ \Pi_j \}$, such that $\Pi_j\ge 0$ and $\sum_j \Pi_j=\openone$.  Each
POVM element represents a single output channel of the measurement
apparatus and the probability of observing the $j$th output is given by
Born rule $p_j=\Tr (\varrho \Pi_j)$.

To proceed further we take advantage of the existence of  a traceless
Hermitian operator basis $\{\Gamma_i\}$ (supplemented with the
identity $\openone$),  satisfying $\Tr (\Gamma_i)= 0$ and
$\Tr (\Gamma_i \Gamma_j) = \delta_{ij}$~\cite{Kimura:2003aa}.  
By decomposing  both $\varrho$ and the POVM elements in
this basis, we get
\begin{equation}
  \label{rhodecomp}
  \varrho= \frac{\openone}{n} + \sum_{i=1}^{n} r_{i} \, \Gamma_i \, ,
  \qquad
  \Pi_j = b_{j} + \sum_{k=1}^{n} a_{jk} \, \Gamma_k\, ,
\end{equation}
where $\{ r_{i}\}$ and  $\{ b_{i}\}$ are real numbers and  $\{
a_{jk}\}$ is a real matrix. Born rule becomes now
\begin{equation}
  \label{eq:2}
  p_{j} = b_{j} + \sum_{k=1}^{n} a_{jk} \,  r_{k} \, , 
\end{equation}
which is again of the form (\ref{eq:tomo}), except for the trivial
addition of a constant vector.

\section{Protocols for quantum detector tomography}

\begin{figure}
  \centering{\includegraphics[height=6cm]{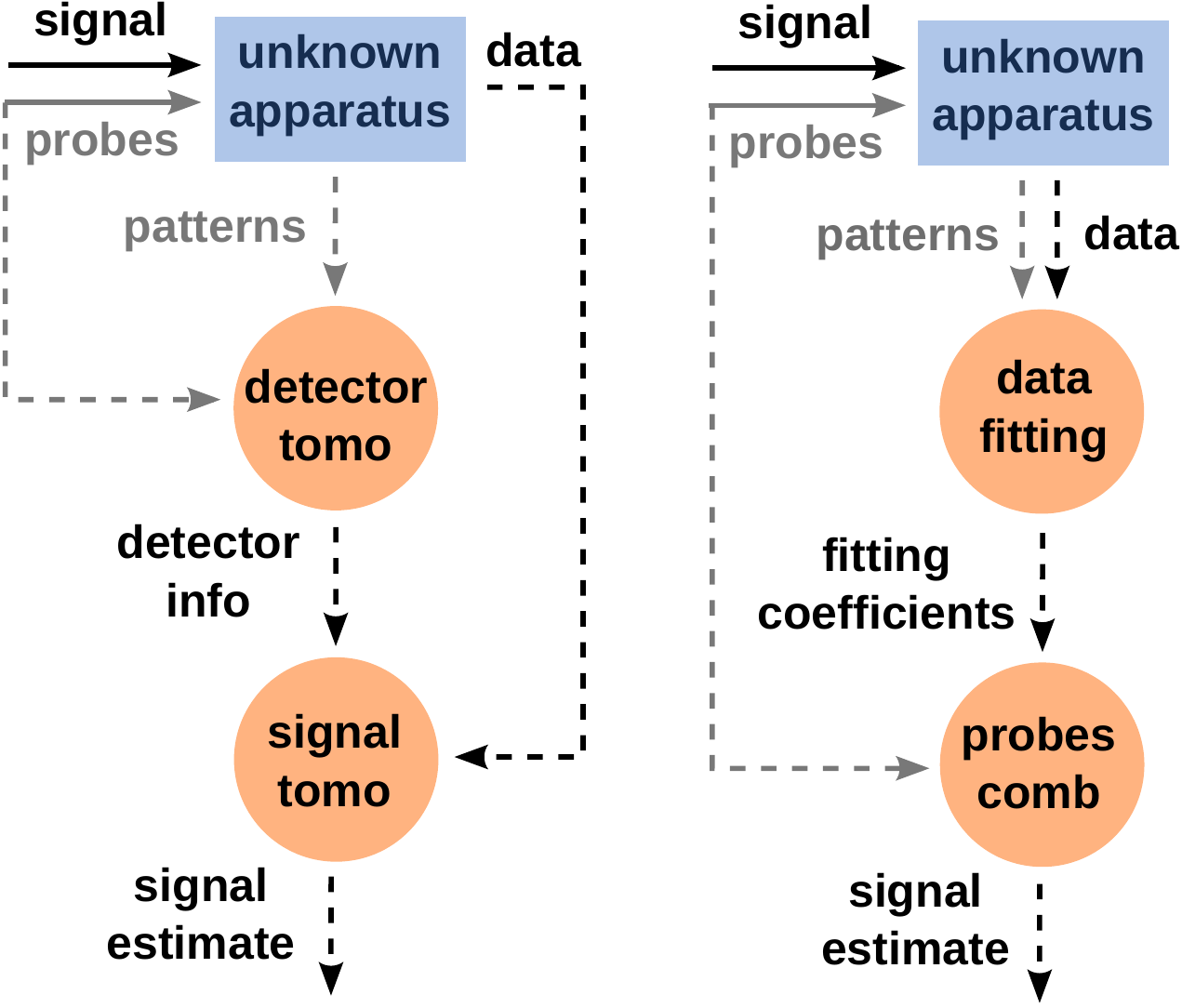}}
  \caption{Information flow in standard tomography (left) and
    data-pattern tomography (right). Blue rectangles and orange
    circles represent the measurement and data processing steps. Solid
   and broken lines (in black for the signal and in grey for the
   probes) denote the propagation of physical systems
    and information, respectively.}
\label{fig:scheme}
\end{figure}

Imagine we wish to determine the state of a quantum system, but the
details of the detection are unknown. As heralded in the
introduction, we have two conceptually different ways of dealing with
this task; they are roughly schematized in figure~\ref{fig:scheme}.  
In both alternatives, a set of known probes is measured and the
corresponding data, called \textit{patterns}, collected. Besides,
additional data for the unknown state we want to estimate is
measured.

%%%%%%%%%%%%%%%%%%%%%%%%%%%%%%%%%%%%
%%%%%%%%%%%%%%%%%%%%%%%%%%%%%%%%%%%%

In the standard approach, one uses the probes and patterns to 
characterize the unknown detector. Afterwards,   the unknown state is
estimated from  data and the measurement matrix previously determined.

In the data-pattern tomography, the best fit of data with a
linear combination of patterns is worked out and the same fit is used
to combine the probes and form the signal estimate. In this way, the
unknown state is expressed directly in terms of probes and their
respective patterns, thus bypassing the (possibly costly) detector
calibration.  We note in passing that similar reasoning is at the
heart of some neural networks for recognition
systems~\cite{Bishop:1995aa}.

We  next examine both protocols from the viewpoint of \textsc{lin}
estimation. 

\subsection{Standard tomography}
\label{sec:standard}

In the first step,  a set of $M$ known probes $\bi{r}_{\alpha}$
($\alpha=1,\ldots, M$) is measured
and their patterns $ \bi{f}_{\alpha }$ collected.  
Arranging probes and patterns columnwise to form the matrices
 $\matriz{R}$ and $\matriz{F}$, of dimensions $m\times M$ and $n
 \times M$, respectively,  we have to solve the problem
\begin{equation}
  \matriz{F} = \matriz{A}  \, \matriz{R},
\end{equation}
where the $m\times n$ matrix $\matriz{A}$ specifies the action of the
detector.  A \textsc{lin}  estimation results in
\begin{equation}
  \label{detectortomo}
  \matriz{A}=\matriz{F} \matriz{R}^{+} .
\end{equation}
The second step is to solve equation~(\ref{inversion}) for the signal
$\mathbf{r}$. Substituting (\ref{detectortomo}), we directly get  
the effective measurement matrix \sugg{inverse} for the standard tomography:
\begin{equation}
  \label{stdrec}
  \matriz{A}_{s} = ( \matriz{F} \matriz{R}^{+})^{+} \, .
\end{equation}
Henceforth, we shall denote this approach by the subscript $s$.

\subsection{Data-pattern tomography}
\label{sec:pattern}

With the same set of probes and patterns we now perform a
data-pattern tomography.  First,  we do a linear
fitting of the data with the  patterns
\begin{equation}
  \label{fit}
  \bi{f}= \matriz{F} \,  \bi{x} \, ,
\end{equation}
where the fitting coefficients $ \bi{x}$ are obtained with
\textsc{lin} inversion; i.e., 
\begin{equation}
  \label{xfit}
   \bi{x}= \matriz{F}^{+} \,  \bi{f} \, .
\end{equation}
Because the linearity of the measurement, the same set of fitting
coefficients can be used to expand the unknown state in known probes;
i.e., $\bi{r}= \matriz{R} \, \bi{x}$.  In this way, we train the
protocol to recognize the probes and their linear combinations.
Finally, using equation~(\ref{xfit}), we get that the effective
measurement matrix \sugg{inverse} for the data-pattern LIN tomography reads
\begin{equation}
  \label{patrec}
  \matriz{A}_{p}= \matriz{R} \matriz{F}^{+} \, ,
\end{equation}
where the subscript $p$ designates this protocol.

\section{Assessing the protocols}

\subsection{Reverse-order law}

A simple glance at equations~(\ref{stdrec}) and (\ref{patrec}) immediately
reveals that both protocols become equivalent provided that
\begin{equation}
  \label{equiv}
  ( \matriz{F} \matriz{R}^{+})^{+}  = \matriz{R} \matriz{F}^{+}  \, .
\end{equation}
This is obviously satisfied by proper matrix inverses, but not always
by pseudoinverses. As sketched in the \ref{sec:MP}, the
identity~(\ref{equiv}) holds true if $\matriz{F}$ and $\matriz{R}$
have full \sugg{column} rank. This leads directly to our main
result: \textsc{lin} standard tomography and \textsc{lin} data-pattern
tomography are equivalent if all probes and all patterns are linearly
independent.

The two protocols will differ if the number of probes and patterns is
sufficiently large $M>\min\{m,n\}$.  Particularly this covers the
following important class of problems, namely
\begin{enumerate}[label=\emph{\alph*})]
\item Informationally complete (IC) settings, where number of probe
  states $M$ is larger than the number of detected channels $m$, which
  equal (or larger) than the number of free parameters $n $
  ($M > m \ge n$.  As will be seen soon, when close to minimal
  tomography $m=n$, data pattern gives better results.
\item IC settings where the number of the output channels $m$ is
  larger (or equal) to number of probe states $M$, which in turn is
  larger than number of free parameters $m \ge M>n$.  Here, the
  standard approach is more efficient.
\item Incomplete measurement, where number of free parameters $n$
  exceeds the number of probe states $M$, which is larger than number
  of detected channels $n\ge  M>m$.  The equality (\ref{equiv})
  does not hold, but such measurements are not tomographically
  complete and  will not be considered here.
\end{enumerate}

\subsection{Performances}

Let us compare the performances of the two tomographic protocols for
IC settings: $M, m \ge n$.  There are two kinds of noise associated
with both protocols: pattern noise and data noise.  We assume that
data noise dominates the reconstruction error.  \sugg{This assumption
  is justified by the very different roles played by the patterns and
  data in experimental tomography: patterns are collected for probe
  states. One would typically use simple states (i.e., easy to
  generate and calibrate) for probes. In general there is no
  restriction on how many copies of probe states can be used for
  probing the measurement setup. On the other hand, the unknown signal
  subject to reconstruction is typically a
  non-trivial/non-classical state and the number of available copies 
  may be limited by the quantum information protocol used. Patterns
  can thus be expected to have better signal to noise ratio than signal data. }

Denoting by $\matriz{A}_{\mathrm{true}}$ the true inversion
matrix \sugg{($\bi{r}=\matriz{A}_{\mathrm{true}} \bi{p}$)}, and by $\Delta \bi{p}$ the data noise
($\bi{f} = \bi{p} + \Delta \bi{p}$), we can straightforwardly obtain
the mean square error
\sugg{
\begin{equation}
\label{eq42}
  e_{\alpha}^2 = 
  \overline{|| \matriz{A}_{\alpha} \bi{f} - \bi{r}||^2} =
  \overline{||\matriz{A}_{\alpha} \bi{f} - \matriz{A}_{\mathrm{true}} \bi{p}||^2}
  \approx 
  \overline{||\matriz{A}_{\alpha} \, \Delta \bi{p}||^2}  =
  \frac{\epsilon^2 \, ||\matriz{A}_{\alpha}||^2}{m} \, . 
\end{equation}
} Here, $\alpha \in \{s, p \}$,
$|| \bi{y} ||^{2} = \Tr (\bi{y}^\ast \bi{y})$ is the Hilbert-Schmidt
norm \sugg{(the symbol $\ast$ denotes here the conjugate transpose)}, and the bar indicates
averaging over all possible realizations of noise with a fixed
strength $||\Delta \bi{p}||=\epsilon$.  \sugg{ To get the last
  inequality, we note that for $||\Delta \bi{p}||=1$ the square norm
  $||\matriz{A}_{\alpha} \, \Delta \bi{p}||^2$ is constrained by the
  minimum and maximum eigenvalues of
  $\matriz{A}_{\alpha}^\ast \matriz{A}_{\alpha}$. Averaging over all
  $\Delta \bi{p}$ yields the aritmetic mean of eigenvalues,
  $\Tr (\matriz{A}_{\alpha}^\ast
  \matriz{A}_{\alpha})/m$. Incidentally, this can be considered as the
  first moment of a real numerical shadow~\cite{Puchala:2012aa} of the
  matrix product $\matriz{A}_{\alpha}^\ast \matriz{A}_{\alpha}$.  For
  $\Delta \bi{p}$ of strength $\epsilon$, we obtain Eq.~(\ref{eq42}).
}

The discussion about
the performance of both protocols thus boils down to a comparison of
the norms of the corresponding inversion matrices, which decides the
average performance of the protocol. Large norms indicate the tendency
to amplify the noise present in the data and, therefore, such
protocols should be avoided.

\subsection{Limiting cases}
\label{sec:limcas}

We next examine two interesting limits that arise in practical
experiments. We shall be making extensive use of a technical
result~\cite{Galperin:1980aa} that provides a unique representation 
for the pseudoinverse of the product of any two matrices; viz,
\begin{equation}
  \label{eq:rep}
  (\matriz{X} \matriz{Y})^{+} = \matriz{Y}^{+} 
  (\matriz{h} + \matriz{g} )\matriz{X}^{+} \, ,
\end{equation}
where $\matriz{h}= (\matriz{X}^{+} \matriz{X}
\matriz{Y}\matriz{Y}^{+})^{+}$ 
is a (skew) projector and $\matriz{g}$ is a matrix with a number of
special properties that the reader can check in the \ref{sec:MP}, as
well as other pertinent details.

\subsubsection{$M> m=n$.}
The measurement thus has the minimal number of outputs
required for IC tomography,  but the set of probes is (highly)
redundant. Let us first assess the standard tomography. Now, 
equation~(\ref{eq:rep}) holds true with $\matriz{g} = 0$; we have then 
\begin{equation}
 \matriz{A}_{s} =(\matriz{F} \matriz{R}^{+})^{+} =   \matriz{R}
 \matriz{h}  \matriz{F}^{+} \, ,
\end{equation}
with $\matriz{h} = (\matriz{F}^{+} \matriz{F}
\matriz{R} \matriz{R}^{+})^{+}$.
We use the singular value decomposition for $\matriz{R}$
as~\cite{Golub:1965aa}   as $\matriz{R} = \matriz{U}_{R} 
\matriz{D}_{R} \matriz{V}_{R}^\ast$ (with $D_{R}$  diagonal,
$\matriz{U}_{R}^\ast \matriz{U}_{R} =
\openone$ and $\matriz{V}^\ast \matriz{V}  = \openone$), and
analogously for  $F$, and $h$. If we call $\mathcal{V}_{Rh}= \matriz{V}_R^\ast
\matriz{U}_ h$ and  $\mathcal{V}_{hF}= \matriz{V}_h^\ast \matriz{V}_F$, we get
\begin{equation}
  ||\matriz{A}_s|| = ||\matriz{S}_R \mathcal{V}_{Rh} \matriz{S}_h
  \mathcal{V}_{hF}\matriz{S}_F^{+}|| \, , 
\end{equation}
where we have introduced  the square representations 
\begin{equation}
  \matriz{S}_{R}=\left (
    \begin{array}{cc}
      \! \matriz{D}_{R} & 0
    \end{array}
  \right ) ,
\qquad 
  \matriz{S}_{h}=\left (
    \begin{array}{cc}
      \matriz{D}_h & 0\\
      0 & 0
    \end{array}
  \right ) \, , 
\qquad
  \matriz{S}_{F}^{+} = \left (
    \begin{array}{c}
      \matriz{D}_F^{+} \\
      0 
    \end{array}
  \right ) \, ,
\end{equation}
all the $\matriz{D}$  blocks being $n\times n$ dimensional. The
unitary matrices $\mathcal{V}_{Rh}$ and $\mathcal{V}_{hF}$ take the
block diagonal form 
\begin{equation}
  \mathcal{V}_{Rh}=\left (
    \begin{array}{cc}
      \matriz{U}_{Rh} & 0\\
      0 & \matriz{W}_{Rh}
    \end{array}
  \right ) \, , 
 \qquad 
  \mathcal{V}_{hF} = \left (
    \begin{array}{cc}
      \matriz{U}_{hF} & 0\\
      0 & \matriz{W}_{hF}
    \end{array}
  \right ),
\end{equation}
with $\matriz{U}_{Rh}, \matriz{U}_{hF}$ and $\matriz{W}_{Rh},
\matriz{W}_{hF}$ being $n\times n$ and 
$(M-n)\times(M-n)$ dimensional, respectively. Consequently,
\begin{equation}
  \label{sand_s}
  ||\matriz{A}_s||=||\matriz{D}_R \matriz{U}_{Rh} \matriz{D}_h
  \matriz{U}_{hF} \matriz{D}_F|| \, , 
\end{equation}
and $||\matriz{U}_{Rh} \matriz{D}_h
\matriz{U}_{hF}||=||\matriz{D}_h|| \ge \sqrt{n}$. 

Repeating the calculation for the pattern tomography, we obtain
\begin{equation}
  \label{sand_p}
  ||\matriz{A}_{p}||=||\matriz{D}_R \matriz{U}_{11} \matriz{D}_F|| \, ,
\end{equation}
where $\matriz{U}_{11}$ is the corresponding matrix element of
$\matriz{V}_{R}^{\ast} \matriz{V}_{F}$, and satisfies
$ |\matriz{U}_{11}| \le\sqrt{n}$.

Of course, the fact that the matrix sandwiched between $\matriz{D}_R$
and $\matriz{D}_F$ in (\ref{sand_s}) has a greater norm than that in
(\ref{sand_p}) does not imply $||\matriz{A}_s|| \ge
||\matriz{A}_p||$. However, with increasing $M$, and hence decreasing
$|\matriz{U}_{11}|$ relative to $||\matriz{D}_h||$, this is likely to
happens and is confirmed in our numerical simulations.  Loosely
speaking, the block diagonal form of $\mathcal{V}_{Rh}$ and
$\mathcal{V}_{hF}$ prevents the scattering of the nonzero singular
values of $\matriz{S}_F^{+}$ into the nullspace of $\matriz{S}_R$,
which tends to increase the norm of the standard inversion matrix,
and, consequently $e_s  \ge e_p$.

We thus conclude that for minimal measurements and highly redundant
set of probes, we expect the \textsc{lin} data-pattern tomography to
outperform the \textsc{lin} standard tomography and the difference
grows with the number of probes $M$.

\subsubsection{$m \ge M > n$.}
Now, the measurement is not minimal and the number of probes does not
exceed the number of measurement outputs. The matrix of patterns
$\matriz{F}$ has full column rank, the corresponding projector
$\matriz{F}^{+}\matriz{F}$ is a unity matrix and, in consequence,
$\matriz{h} = (\matriz{R}^{+}\matriz{R})^{+}= \matriz{R}^{+}
\matriz{R}$:  
\begin{equation}
  \label{case2}
  \matriz{A}_s= \matriz{R} (\matriz{R}^{+} \matriz{R} +\matriz{g})
  \matriz{F}^{+} \, . 
\end{equation}
By the properties of $\matriz{g}$, and more concretely
(\ref{minnorm}), we have
\begin{equation}
  \label{minnorm2}
  ||\matriz{A}_s||=||\matriz{R} (R^{+}R+g)F^{+}||
\le ||R F^{+}||=||\matriz{A}_p||
\end{equation}
and $e_s \le e_p$. Hence, for highly complex measurements,  we expect
the standard detector tomography to be more efficient than the
data-pattern approach.  

To sump up, choosing a fixed $m \ge n$ and increasing
the number of probes from $M=n$, the data pattern tomography,
initially the worse technique, becomes superior for large probe
numbers $M\gg n$.

\section{Examples}

%%%%%%%%%%%%%%%%%%%%%%%%%%%%%%%%%%
\begin{figure}[b]
  \centering{\includegraphics[width=0.48 \columnwidth]{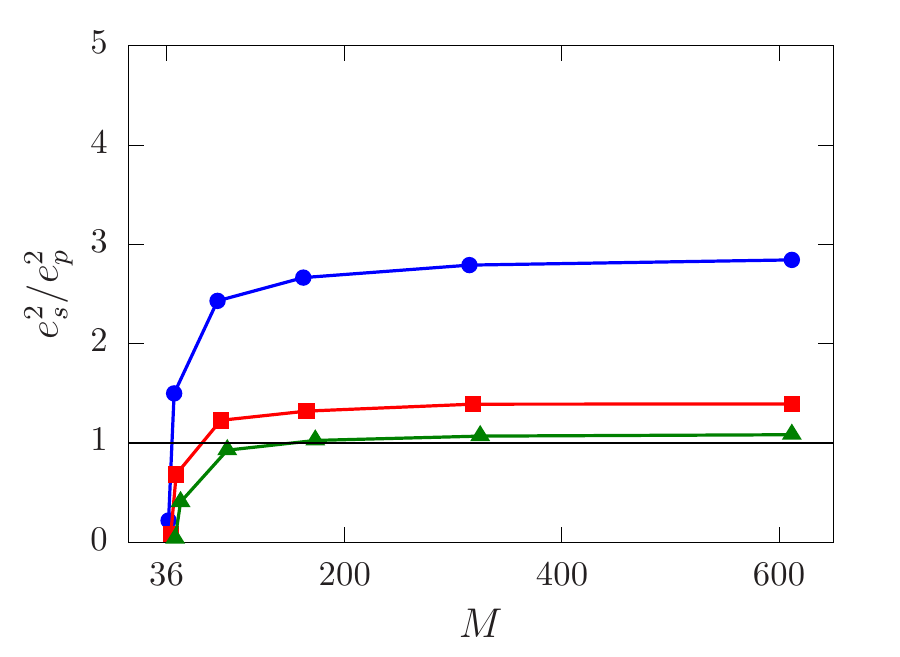}
    \includegraphics[width=0.48 \columnwidth]{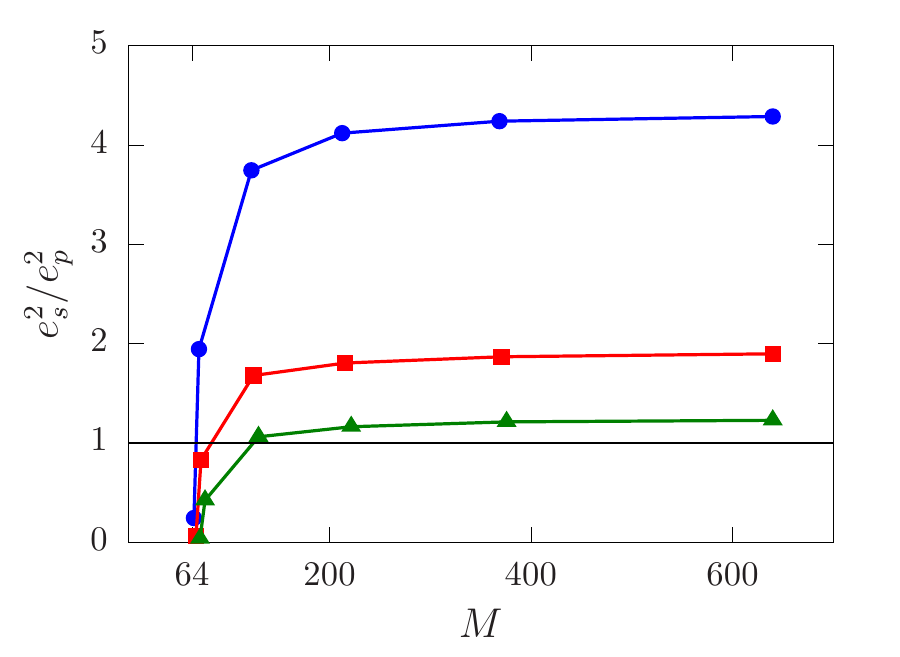}}
  \caption{Performance ratios in dimension $d=6$ ($n=35$, left) and
    $d=8$ ($n=63$, right). Random square-root measurements with
    $m=n+3$ (blue circles), $m=n+5$ (red squares), and $m=n+9$ (green
    triangles) outputs are simulated.}
  \label{figsimM}
\end{figure}
%%%%%%%%%%%%%%%%%%%%%%%%%%%%%%%%%%%%

To confirm our conclusions we have performed simulations of
quantum state tomography with random square-root measurements
\begin{equation}
  \label{squareroot}
  \Pi_j = G^{-1/2} |\phi_j\rangle\langle \phi_j| G^{-1/2} \, , 
  \qquad 
   G=\sum_j  |\phi_j\rangle\langle \phi_j| \, , 
\end{equation}
where $|\phi_j\rangle$ are randomly generated Haar-distributed pure
states (aka random states) and $j= 1, \ldots, m$. This scheme has been
proposed as  \emph{a pretty good}' measurement for distinguishing
quantum states~\cite{Hausladen:1994aa} and is known to be
optimal~\cite{Dalla-Pozza:2015aa}.

For each measurement we generate a set of $M$ random probes and the
corresponding patterns. The simulated data is obtained for a set of
10000 random true states and the mean square errors are
computed from the estimates. Finally, to quantify the performance, we
average $e_s^2$ and $e_p^2$ over several hundred random square-root
measurements.  The system dimension $d$ is fixed to be 6 or 8 (which
means $n= d^{2}-1$ is 35 and 63, respectively) and the noise was
simulated by adding Gaussian noise to the theoretically calculated
detection probabilities. The noise-to-signal ratio was set $0.03$ and
$0.06$ for the patterns and data, respectively.

The first set of simulations in figure~\ref{figsimM} shows the
dependence of the performance ratio $e_s^2/e_p^2$ on the number of
probes for three different  measurement outputs $m$.  In a
remarkable agreement with the theory, the data-pattern approach performs
poorly when $m=M$,  but considerably improves when the number of
probes increases; and this is more pronounced for higher
dimensions. 

%%%%%%%%%%%%%%%%%%%%%%%%%%%%%%%
\begin{figure}
  \centering{\includegraphics[width=0.48 \columnwidth]{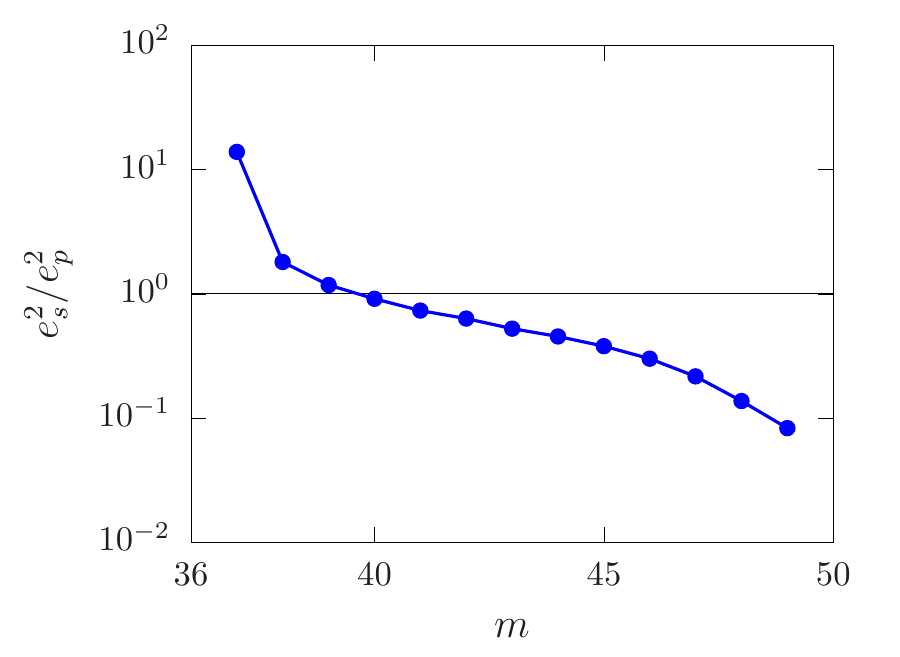}
    \includegraphics[width=0.48\columnwidth]{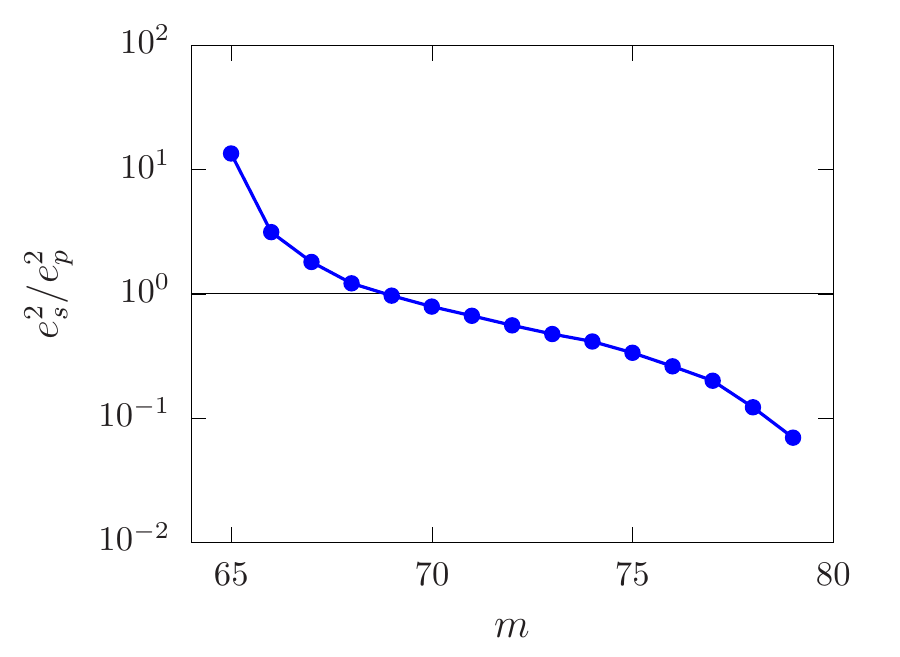}}
  \caption{ Performance ratios for random square-root measurements in
    dimension $d=6$ with $M=50$ probes (left), and dimenson
    $d=8$ and $M=80$ probes (right).}
  \label{figsimN}
\end{figure}
%%%%%%%%%%%%%%%%%%%%%%%%%%%%%%%%

In figure~\ref{figsimN}, we plot the same ratio $e_s^2/e_p^2$, but now
keeping $M$ fixed and varying $m$. The data-pattern approach is better
for minimal and nearly minimal schemes $m \approx n$, whereas the
standard tomography is recommended for highly complex measurements
$m \gg n$.

%%%%%%%%%%%%%%%%%%%%%%%%%%
\begin{figure}[b]
  \centering{\includegraphics[width=0.48 \columnwidth]{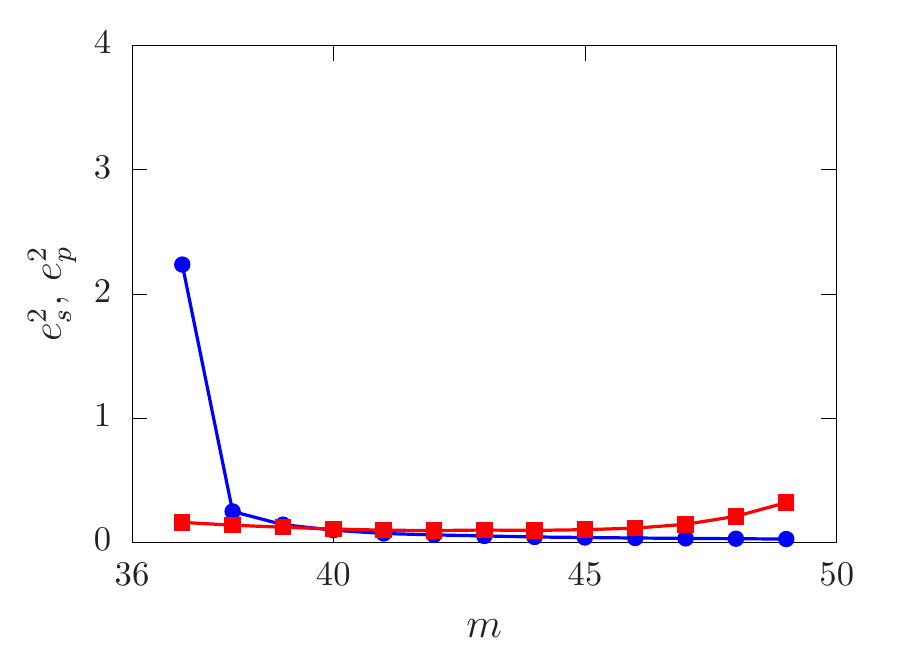}
    \includegraphics[width=0.48 \columnwidth]{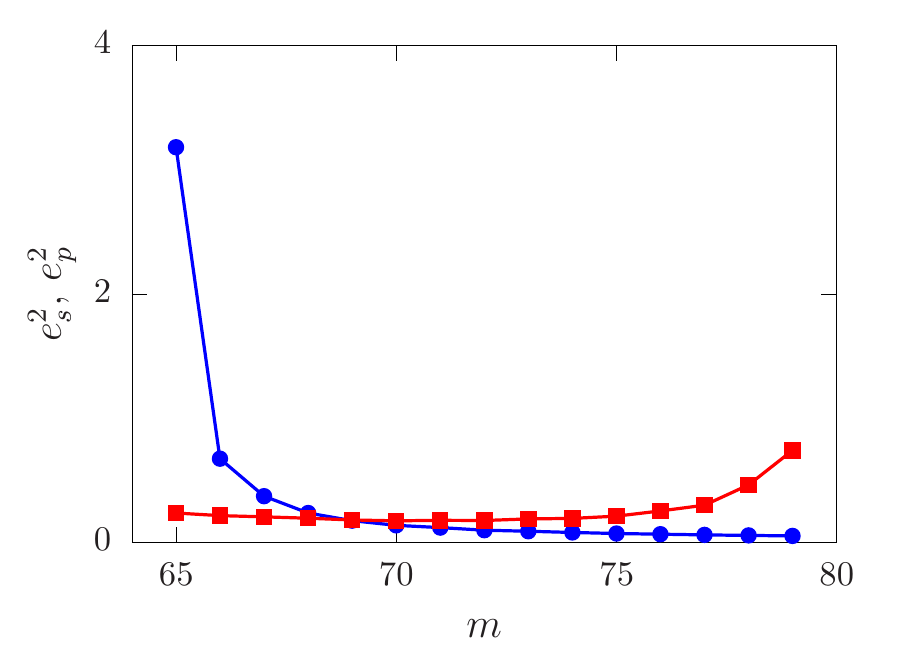}}
 \caption{Mean square errors for random square-root measurements with 
the standard (blue circles) and data pattern
(red squares) protocols  in dimension $d=6$ ($n=35$ and $M=50$, left),
and $d=8$ ($n=63$ and $M=80$, right).  
\label{figsim2N}}
\end{figure}
%%%%%%%%%%%%%%%%%%%%%%%%%%

Figure~\ref{figsim2N} shows the mean square errors for the two
protocols. With a fixed number of probes, the reconstruction error of
the standard scheme decreases with increasing $m$: more
measurement outputs make the overall reconstruction error smaller. For
the data-pattern approach, this tendency is much the opposite.

We also simulated realistic homodyne detection. For this end,  we
have used  coherent states as probes, as they are easy to
generate. In our simulations we use of $M=100$ randomly generated
coherent states with amplitudes $|\alpha|< 0.8$. The states are
calibrated and their amplitudes are known to the experimenter. The
true unknown state to be reconstructed is chosen to be the following
coherent superposition (in the Fock basis )
\begin{equation}
  \label{rhotrue}
  |\Psi_{\mathrm{true}} \rangle= \sqrt{0.1}  \, |0\rangle 
 +\sqrt{0.2} \, |1\rangle  +\sqrt{0.3} \, |2\rangle \, .
\end{equation}
We set the quantum efficiency of the homodyne detection to $80\%$ and
generate $m$ random quadrature measurements of the signal and probe
states. The data-pattern tomography and standard tomography are
used to reconstruct the unknown signal in the subspace spanned by the
Fock states $|0\rangle,|1\rangle,\ldots, |5\rangle$, so that $n=35$.

%%%%%%%%%%%%%%%%%%%%%%%
\begin{figure}
  \centering{\includegraphics[width=0.52 \columnwidth]{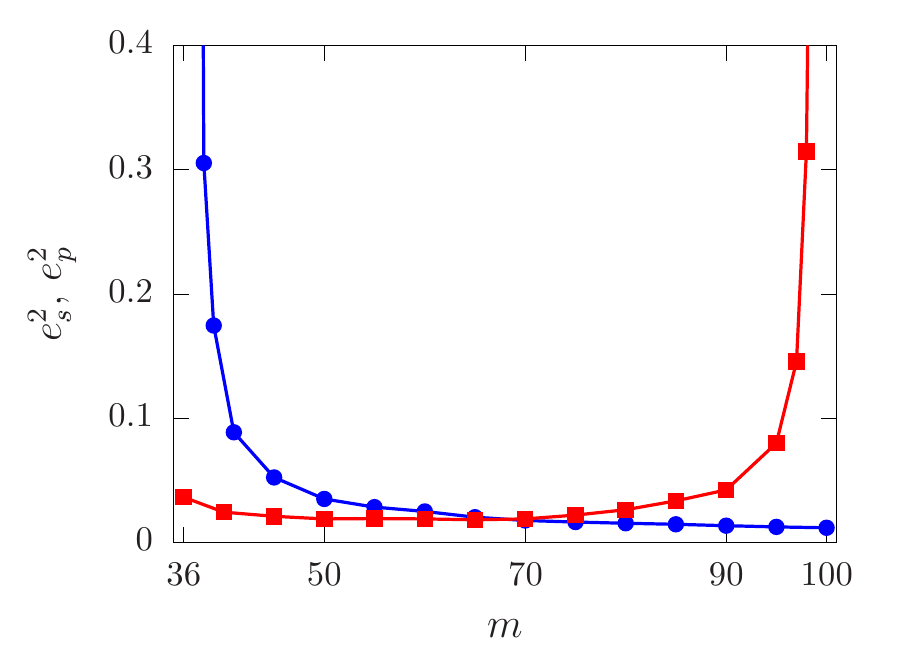}}
\caption{Mean square errors of standard (blue circles) and
  data-pattern (red squares) protocols  in the Fock subspace of
  dimension six ($n=35$). State averaged errors are shown for a
  randomly generated inefficient homodyne measurement with $\eta=0.8$,
  and $M=100$. 
  \label{figcomparhomo}}
\end{figure}
%%%%%%%%%%%%%%%%%%%%%%%

\begin{figure}
 \centering{ \includegraphics[width=0.42\columnwidth]{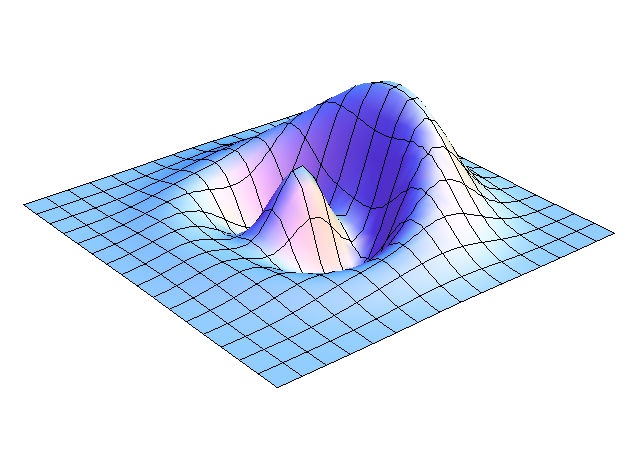}
 \includegraphics[width=0.42\columnwidth]{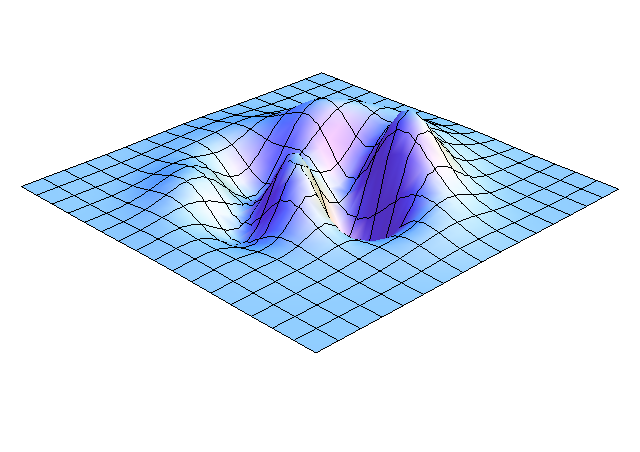}}
\caption{Typical reconstructions by data-pattern tomography   with
  $m=n=35$ (left) and  $m=M=100$ (right). In the minimal IC setting
  the reconstruction is very good (in fact, almost identical to the
  Wigner function for the true state, with strong negativity
  witnessing the nonclassical character), while it is very bad in the
  other opposite limit. The reverse behavior, with almost the same
  figures interchanged, is observed for the standard approach,
  although the figures are not reproduced here for simplicity.} 
  \label{figwig}
\end{figure}

As one can observe in figure~\ref{figcomparhomo}, the mean square
errors display two prominent resonant-like peaks at $m=n$ and $m =M$,
where one protocol clearly outperforms the other.
 This can be illustrated further with the Wigner representations of the
measured signal obtained in those two extreme regimes, shown in 
figure~\ref{figwig}. For example, for the minimal IC setting $m=n$, the
repeated data-pattern reconstructions are all very good, while the
standard tomography delivers results ranging from reasonably good to
very bad.

%%%%%%%%%%%%%%%%%%%%%%%%%%%%%%
%%%%%%%%%%%%%%%%%%%%%%%%%%%

\section{Conclusions}

In summary, we have found substantial differences in the performance
of the two protocols traditionally used for estimating the state of a
physical system from unknown measurements. This is important from
a practical point of view, as obtaining good results strongly depends on
choosing the correct estimation strategy.  In particular, the
data-pattern approach yields better results for minimal measurement
settings, whereas the standard approach is better suited to handle
complex overcomplete measurements.  \sugg{These differences can be 
 ascribed to different order in which the information 
available is processed to get the final signal estimate. Since we deal
with matrices, the order of utilizing information about probes and
patterns does matter.} 

Although our theory and simulations have been worked out solely for
the simple linear estimator, we expect to see similar effects should a
more sophisticated nonlinear inference, such as the maximum-likelihood
estimation, be adopted. This is the subject of further investigations.

\ack 
We acknowledge financial support from the Technology Agency of
the Czech Republic (Grant TE01020229), the Grant Agency of the Czech
Republic (Grant No. 15-03194S), the IGA Project of the Palack{\'y}
University (Grant No. IGA PrF 2016-005) and the Spanish MINECO (Grant
FIS2015-67963-P).

\appendix

\section{The Moore-Penrose pseudoinverse}
\label{sec:MP}

In these appendix we give a brief account of  the Moore-Penrose
pseudoinverse. The interested reader can find more details in the
comprehensive references \cite{Ben-Israel:1977aa} and
\cite{Campbell:1991aa}. 

For any $m \times n$ matrix $\matriz{X}$, its pseudoinverse
$\matriz{X}^{+}$ must satisfy all the following
conditions~\cite{Penrose:1955aa}
\begin{enumerate}[label=C\arabic{enumi}.]
\item $\matriz{X} \matriz{X}^{+} \matriz{X} = \matriz{X}$
\item $\matriz{X}^{+} \matriz{X} \matriz{X}^{+} = \matriz{X}^{+}$
\item ($\matriz{X} \matriz{X}^{+})^{\ast} = \matriz{X} \matriz{X}^{+}$
\item ($\matriz{X}^{+} \matriz{X})^{\ast} = \matriz{X}^{+} \matriz{X}$
\end{enumerate}
Here $\ast$ denotes the conjugate transpose. This provides a quick
checkable criterion; that is, given a matrix $G$ that purports to be
the pseudoinverse of $X$, one need simply verify the four conditions
C1--C4 above. This verification is often relatively straightforward.
We stress that the pseudoinverse always exists and is unique.

Alternatively, $\matriz{X}^{+}$ can be defined  as the limit
\begin{equation}
  \label{eq:1}
  \matriz{X}^{+} = \lim_{\delta \rightarrow 0} (
 \matriz{X}^{\ast}  \matriz{X} + 
\delta \openone)^{-1}  \matriz{X}^{\ast} =
 \lim_{\delta \rightarrow 0} 
\matriz{X}^{\ast} (\matriz{X}  \matriz{X}^{\ast} + 
\delta \openone)^{-1}\, .
\end{equation}

We next quote a few useful properties:
\begin{enumerate}[label=\arabic{enumi}.]
\item $\matriz{X}^{+} =  (\matriz{X}^{\ast}  \matriz{X})^{+}
  \matriz{X}^{\ast} =    \matriz{X}^{\ast} (\matriz{X}
  \matriz{X}^{\ast})^{+}$
\item $(\matriz{X}^{\ast})^{+} = (\matriz{X}^{+})^{\ast}$  
\item $(\matriz{X}^{+})^{+}  = \matriz{X}$
\item $(\matriz{X}^{\ast} \matriz{X})^{+} = \matriz{X}^{+}
  (\matriz{X}^{\ast})^{+}, \quad (\matriz{X}  \matriz{X}^{\ast})^{+} =
  (\matriz{X}^{\ast})^{+}  \matriz{X}^{+}$ 
\item $ \mathcal{R} ( \matriz{X}^{+}) = \mathcal{R} (\matriz{X}^{+} \matriz{X})
  = \mathcal{R} (\matriz{X}^{t}),
  \quad 
\mathcal{N} (\matriz{X}^{+}) = \mathcal{N} ( \matriz{X}
\matriz{X}^{+})  =
\mathcal{N}  (\matriz{X}^{t}) $
\end{enumerate}
where $\mathcal{R}$ denotes the range and $\mathcal{N}$ the null
subspace of the corresponding matrix, \sugg{and the superscript $t$
  stands for the transpose.}

If $\matriz{X}$ is $m\times n$ and $\matriz{Y}$ is $n\times p$ and
either: $ \matriz{X}$ has orthonormal columns, or $\matriz{Y}$ has
orthonormal rows, or $\matriz{X}$ has full column rank and
$\matriz{Y}$ has full row rank, or $\matriz{Y}= \matriz{X}^{\ast}$,
then
\begin{equation}
  \label{eq:3}
(\matriz{X} \matriz{Y} )^{+} = \matriz{Y}^{+} \matriz{X}^{+} \, .
\end{equation}
In the general case, however, the analytical structure of the Moore-Penrose
pseudoinverse of the product of any two matrices is more complex. It
has been fully specified  by Galperin and
Waksman~\cite{Galperin:1980aa}, and the following theorem
has been extensively used in our paper:
\begin{thm}
Let $\matriz{X}$ and $\matriz{Y}$ be two arbitrary matrices, and $h = (\matriz{X}^{+}
\matriz{X} \matriz{Y}\matriz{Y}^{+})^{+}$. Then there exists a unique 
matrix $\matriz{g}$ such that
 \begin{equation}
  \label{theorem}
  (\matriz{X} \matriz{Y} )^{+}= \matriz{Y}^{+} 
  (h+g) \matriz{X}^{+} 
\quad \&  \quad
\matriz{Y} \matriz{Y}^{+} g \matriz{X}^{+} \matriz{X} =g \,.
\end{equation}
It holds that $g \perp h$ and of all $z$ satisfying 
$\matriz{Y}\matriz{Y}^{-}z \matriz{X}^{+} \matriz{X}=z$ and
$\matriz{X} z \matriz{Y}=0$, $g$ makes $||\matriz{Y}^{+} (h+z)
\matriz{X}^{+}||$ minimum.  In particular, this
means that
\begin{equation}
  \label{minnorm}
  || \matriz{Y}^{+}(h+g) \matriz{X}^{+}|| \le 
  ||\matriz{Y}^{+} h  \matriz{X}^{+}|| \, .
\end{equation}
\end{thm}
The projector $h$ has many interesting properties.  It holds true that
\begin{equation}
\begin{array}{l}
  \mathcal{R} (h) = \mathcal{R} ( \matriz{Y} ) \cap [ \mathcal{R} (\matriz{X}^{\ast}) +
  \mathcal{N} (\matriz{Y}^{\ast})] \, ,  \\
 \mathcal{N} (h) = \mathcal{R} (\matriz{X}^{\ast} ) \cap \mathcal{N} (\matriz{Y}^{\ast})
  + \mathcal{N} (\matriz{X}) \, . 
\end{array}
\end{equation}
These properties have played a crucial role in our estimates in
section~\ref{sec:limcas}. 

\newpage

%\bibliographystyle{iopart-num} 
%\bibliography{Tomo}
%\input{patLIN_v4.bbl}

\providecommand{\newblock}{}

\end{document}